\documentclass[12pt,epsf]{article}
\usepackage{graphicx, amsmath}

\begin{document}
\sloppy
\begin{flushright}{SIT-HEP/TM-50}
\end{flushright}
\vskip 1.5 truecm
\centerline{\large{\bf Delta-N formalism for the evolution of the
curvature perturbations}}
\centerline{\large{\bf in generalized multi-field inflation}}
\vskip .75 truecm
\centerline{\bf Tomohiro Matsuda\footnote{matsuda@sit.ac.jp}}
\vskip .4 truecm
\centerline {\it Laboratory of Physics, Saitama Institute of Technology,}
\centerline {\it Fusaiji, Okabe-machi, Saitama 369-0293, 
Japan}
\vskip 1. truecm

\makeatletter
\@addtoreset{equation}{section}
\def\theequation{\thesection.\arabic{equation}}
\makeatother
\vskip 1. truecm

\begin{abstract}
\hspace*{\parindent}
The $\delta N$ formalism is considered to calculate the 
evolution of the curvature perturbation in generalized
 multi-field inflation models. 
The result is consistent with the usual calculation of the standard
 kinetic term. 
For the calculation of the generalized kinetic term, we improved the
 definition of the adiabatic field.
Our calculation improves the usual calculation of $\dot{\cal R}$ based
 on the field equations and the perturbations, giving a very simple and
 intuitive argument for the evolution equations in terms of the
 perturbations of the inflaton velocity.
Significance of non-equilibrium corrections are also discussed, which is
 caused by the small-scale (decaying) inhomogeneities.
This formalism based on the modulated inflation scenario (i.e.,
 calculation based on the perturbations related to the inflaton
 velocity) provides a 
 powerful tool for investigating the signature of moduli that may appear
 in string theory. 
\end{abstract}

\newpage
\section{Introduction}
Inflation has become a major paradigm for explaining the very early
Universe that is consistent with the observations, and current
observations of the temperature anisotropy of the Cosmic Microwave
Background (CMB) support the scale-invariant and Gaussian spectrum that
is expected from the standard inflation scenario. 
However, there are some anomalies in the spectrum, such as a small
departure from the exact scale-invariance or a certain non-Gaussian
character\cite{Bartolo-text}, which are expected to reveal the dynamics
of the fields that are responsible for the inflation. 
An obvious example is the observation of a small shift in the spectrum
index $n-1\ne 0$, which suggests that there is a small departure from
the scale-invariant evolution and which has been used to constrain the
inflation potential  \cite{EU-book}.
More recently, it has been claimed that observations may support a
significant non-Gaussian character in the spectrum
\cite{NG-obs}.
For example, with regard to the generation of the anomaly that captures
the inflation dynamics, there are models for inflation in which the
spectrum is generated
(1) during inflation \cite{Step-inflation, Nonlocal_DBI_inflation,
Modulated1, Modulated-Kin, hybrid-matsuda}, (2) at the 
end of inflation \cite{End-Modulated, End-multi, End-multi-mat},  (3)
after inflation by preheating and reheating
\cite{IH-PR,IH-string,IH-R,Preheat-ng, kin-NG-matsuda}   
 or (4) by the curvatons \cite{curvaton-paper, curvaton-dynamics,
Mazumdar-curvaton, matsuda_curvaton}.
In addition to these scenarios, an inhomogeneous phase
transition\cite{inhomogeneous-pt} may play crucial role in generating
cosmological perturbations.
In this paper, 
the first possibility is considered; the correction during inflation
causes a significant anomaly in the spectrum. 
We have a special interest
in the effects of massless excitations during inflation, which may
capture the extra-dimensional structure of the
Universe.\cite{Kofman-modulated}  

Consider multi-field inflation with kinetic terms described by a metric
$G^{IJ} (\phi^K)$ in field space and $n$ scalar fields.
The action of the model can basically be described by 
\begin{equation}
\label{action-1}
S=\int d^4 x \sqrt{-g}\left[
\frac{R}{16\pi G}+P(X, \phi^I)
\right],
\end{equation}
where $X$ is given by
\begin{equation}
X\equiv -\frac{1}{2}G_{IJ}\partial_\mu \phi^I\partial^{\mu}\phi^J.
\end{equation}
For simplicity two-field inflation is considered in this paper, in which
the adiabatic field $\sigma$ and the entropy field $s$ appear.
Here the adiabatic and entropy fields are defined by
$\dot{\sigma}^2\equiv \sum_I (\dot{\phi^I})^2=2X$ and $s$
($\dot{s}\equiv 0$). 
Using the kinetic part $K$ and the potential $V$, $P(X,\phi^I)$ is
expressed as $P(X,\phi^I)=K(X,\phi^I)-V(\phi^I)$.

Before discussing the details of the calculational method, which is
based on the modulated inflaton velocity (``modulated
inflation'' in ref.\cite{Modulated1}),  it would be useful to
show why in the $\delta N$ formalism  
the modulated inflaton velocity can be used for the evolution equation
of the curvature perturbation, instead of using the
conventional non-adiabatic pressure perturbation or the time-derivative
of the comoving curvature perturbation.

Following the traditional calculation \cite{EU-book}, the spectrum of the
curvature perturbation ${\cal P}_{\cal R}(k)$ for the (adiabatic)
inflaton field $\sigma$ is given by
\begin{equation}
\label{bound-ini}
{\cal P}_{\cal R}(k) = \left(\frac{H}{\dot{\sigma}}\right)^2
\left(\frac{H}{2\pi}\right)^2,
\end{equation}
where $H$ is the Hubble parameter of the Universe and the right-hand
side is evaluated at the epoch of the horizon exit $k=aH$.
Here $a$ is the cosmic-scale factor.
In the above equation, the comoving curvature perturbation ${\cal R}$
is considered, 
which can be identified with the curvature perturbation
on uniform-density hypersurfaces $\zeta$ (${\cal R}\simeq -\zeta$) 
 by studying the evolution of $\zeta$ at large
scales. 
The gauge-invariant combinations for the curvature
perturbation can be constructed as follows:
\begin{eqnarray}
\label{zeta-org}
\zeta &=&-\psi -H\frac{\delta \rho}{\dot{\rho}}\nonumber\\
{\cal R}&=& \psi -H\frac{\delta q}{\rho+p},
\end{eqnarray}
where $\delta q =-\dot{\sigma}\delta
\sigma$  is the momentum perturbation satisfying
\begin{equation}
\epsilon_m=\delta \rho-3H \delta q,
\end{equation}
where $\epsilon_m$ is the perturbation of the comoving density.
Linear scalar perturbations of a
Friedman-Robertson-Walker(FRW) background were considered:
\begin{equation}
ds^2=-(1+2A)dt^2 + 2a^2(t)\nabla_i B dx^i dt +a^2(t)
[(1-2\psi)\gamma_{ij}+2\nabla_i\nabla_j E]dx^i dx^j.
\end{equation}
Here $\rho$ and $p$ denote the energy density and the pressure 
during inflation.

 Note that in the traditional argument the evolution of the curvature
 perturbation at large scales
is calculated to be given by the non-adiabatic pressure perturbation
$\delta p_{nad}$; 
\begin{equation}
\label{111}
\dot{\zeta}\simeq -H\frac{\delta p_{nad}}{\rho+p},
\end{equation}
where $\zeta$ and ${\cal R}$ coincide ($\zeta\simeq-{\cal R}$)
at large scales and 
$\delta p_{nad}\equiv\left[\delta p
-\frac{\dot{p}}{\dot{\rho}}\delta \rho\right]$
is a gauge-invariant perturbation.

In fact, from Eq.(\ref{zeta-org}), it is found that
\begin{equation}
\label{t-der-zeta}
\dot{\zeta} =-\dot{\psi}-\frac{d}{dt}\left[
H\frac{\delta \rho}{\dot{\rho}}\right],
\end{equation}
and the equations for the local conservation of the
energy momentum lead to\cite{A-NEW}
\begin{equation}
\label{enemoeq}
\dot{\delta \rho}=-3H(\delta \rho+\delta p)
+(\rho+p)\left[3\dot{\psi}-\nabla^2 (\sigma + v + B)\right],
\end{equation}
where the scalar describing the shear is
\begin{equation}
\sigma = \dot{E}-B
\end{equation}
and $\nabla ^i v$ is the perturbed 3-velocity of the fluid.
Eq. (\ref{enemoeq}) gives the equation for $\dot{\psi}$, which can be
used to derive 
\begin{eqnarray}
\label{t-der-zeta2}
\dot{\zeta} &\simeq&
-\frac{\dot{\delta \rho}+3H(\delta \rho + \delta p)}{3(\rho+p)}
+\frac{d}{dt}\left[H\frac{\delta \rho}{3H(\rho+p)}\right]\nonumber\\
&=& -\frac{H}{\rho+p}
\left[\delta p-\frac{\dot{p}}{\dot{\rho}}\delta \rho \right]\nonumber\\
&=& -\frac{H}{\rho+p} \delta p_{nad} 
\end{eqnarray}
where $\nabla^2 (\sigma + v + B)$ is
neglected.

Besides $\dot{\zeta}$ defined above, it is useful to define the
perturbed expansion rate with respect to the 
coordinate time\footnote{See Ref. \cite{A-NEW, delta-N-ini}
 for more details on the definitions.}
\begin{equation}
\delta \tilde{\theta}\equiv -3\dot{\psi}+\nabla^2\sigma,
\end{equation}
which leads to\footnote{$\frac{d}{dt}\left[H\frac{\delta\rho}{\dot{\rho}}\right]$  
can be identified with the shear when the perturbation $\delta \rho$
is defined on spatially flat slicing and it
satisfies the adiabatic condition\cite{A-NEW}. 
The shear at large scales is neglected in the above equation, but terms
 related to non-adiabatic perturbations are not disregarded.} 
\begin{eqnarray}
\label{del-n-t-2}
\frac{d}{dt}
\delta N &\equiv& \frac{1}{3}\delta \tilde\theta \nonumber\\
&\simeq& -\dot{\psi}\nonumber\\
&=& \dot{\zeta} +
\frac{d}{dt}\left(H\frac{\delta \rho}{\dot{\rho}}\right)
\nonumber\\
&=&-\frac{H}{\rho+p} \delta p_{nad} +
\frac{d}{dt}\left(H\frac{\delta \rho}{\dot{\rho}}\right).
\end{eqnarray}
Following the conventional definition of the $\delta N$ formalism,
{\bf we choose the gauge whose slicing is flat at $t_{ini}$ and uniform
density at 
$t$}.
Using $\zeta =-\psi$ on the specific choice of slice at
$t$, the $\delta N$ formula is given by
\begin{equation}
\zeta = \frac{1}{3}\int^t_{t_{ini}}\delta \tilde{\theta}dt =\delta N,
\end{equation}
which shows that Eq. (\ref{t-der-zeta2}) and Eq.(\ref{del-n-t-2}) are
consistent, since the equation is for the curvature perturbation $\zeta$
on uniform density slice at $t$.
Here the equation for the perturbed expansion rate $\delta
\tilde{\theta}$ for the $\delta N$ formula is practically valid in any
gauge and slicing, but the relation between the curvature perturbation
and $\delta N$ is defined for the specific choice of slice at
$t$. 

%del 1-start%
%del 1-end%
We also define $\dot{\zeta}_{N}$ in terms of the $\delta N$ formalism 
defined for the uniform density hypersurfaces.
From Eqs. (\ref{del-n-t-2}) and (\ref{enemoeq})
\begin{eqnarray}
\label{orig}
\dot{\zeta}_{N} &\equiv& \frac{d}{dt}\delta N
 = -\dot{\psi}\nonumber\\
&\simeq&-
H\frac{\delta (\rho+p)}{(\rho+p)}-H\frac{\dot{\delta \rho}}{3(\rho+p)}
\nonumber\\
&\simeq&-
H\frac{\delta (\rho+p)}{(\rho+p)}\nonumber\\
&=& -H\frac{\delta(\dot{\sigma}^2)}{\dot{\sigma}^2},
\end{eqnarray}
where the adiabatic field $\sigma$ is defined so that the action has
the standard kinetic term.\footnote{Due to the approximations that have been considered in
deriving the equation, the result is not exact with regard to the shear
perturbations that accompany $k^2/a^2$ factor.}
Here $\dot{\delta \rho}$ has been disregarded in the uniform
density gauge.
The basic formula of the evolution of $\delta N$ is valid for any
gauge and slicing, but the relation between $\dot{\zeta}$ and 
$\dot{\delta N}$ is defined in the specific slicing.
As a result, in terms of the $\delta N$ formalism, the evolution of the
curvature perturbation can be
 explained using the perturbations related to the 
inflaton velocity ($\delta(\dot{\sigma}^2)$)\cite{Modulated1}.
In this paper, using the $\delta N$ formalism defined for the uniform
density hypersurfaces at $t$, a very simple 
method for calculating the evolution of the curvature perturbation
is discussed.

\section{Standard kinetic term}

\underline{\bf The model}

In order to explain the validity of the calculational method based on
the modulated inflation 
scenario, and to explain how the method makes calculation very easy
and clear,
first the simplest model of
multi-field inflation is considered, which is characterized by 
\begin{equation}
\label{simplest-toy}
P=X-V,
\end{equation}
where the metric $G^{IJ}$ in the definition of $X$ is equivalent to the
unit matrix (i.e. consider $G_{IJ}=\delta_{IJ}$) and
$V(\phi^1,...,\phi^n)$ represents the scalar potential.  
The field equation derived from Eq. (\ref{action-1}) with
the condition (\ref{simplest-toy}) is
\begin{equation}
\ddot{\phi}^I+3H\dot{\phi}^I+V_I=0,
\end{equation}
where $V_I\equiv \frac{\partial V}{\partial \phi^I}$, and the Hubble rate
$H$ is determined by the Friedman equation:
\begin{equation}
H^2 \equiv \frac{1}{3M_p^2}\left[
\frac{1}{2}\sum_I \left(\dot{\phi}^I\right)^2+V\right],
\end{equation}
where $M_p$ is the reduced Planck mass.
In a spatially flat FLRW space-time with metric given by
\begin{equation}
ds^2=-dt^2+a^2(t)d\vec{x}^2,
\end{equation}
 the scalar fields are homogeneous.
In order to study cosmological fluctuations, the scalar perturbations of
the metric are: 
\begin{equation}
ds^2 = -(1+2 A)dt^2 + 2aB_idx^idt + a^2\left[(1-2\psi)\delta_{ij}+2E_{ij}
\right]dx^i dx^j,
\end{equation}
which leads to energy and momentum perturbations given in terms of
the scalar field perturbations;
\begin{eqnarray}
\delta \rho &=& \delta X +\delta V\\
3H \delta q &=& -3H\sum_I \dot{\phi}^I\delta \phi^I\nonumber\\
&=& \sum_I \left[\ddot{\phi^I}+V_I\right]\delta \phi^I,
\end{eqnarray}
where the last equation is obtained by using the field equation.
Combining these equations, the comoving density perturbation
is found to be given by
\begin{eqnarray}
\epsilon_m &\equiv& \delta \rho-3H\delta q\nonumber\\
&=&\delta X + \left(\delta V -\sum_I V_I \delta \phi^I\right)
-\sum_I \ddot{\phi}^I\delta \phi^I\nonumber\\
 &\equiv&\delta X + \delta^{(2)} V
-\sum_I \ddot{\phi}^I\delta \phi^I,
\end{eqnarray}
which is a gauge-invariant quantity.
Here $\delta^{(2)} V$ denotes higher order corrections with respect to the
field perturbations.
The comoving density perturbation satisfies the evolution equation 
\begin{equation}
\label{decay-comv}
\epsilon_m =-\frac{1}{4\pi G}\frac{k^2}{a^2}\Psi,
\end{equation}
where $\Psi$ is related to the shear perturbation.

\underline{\bf Calculation}

The adiabatic and entropy fields are defined by
$\dot{\sigma}^2\equiv \sum_I (\dot{\phi^I})^2=2X$ and $s$
($\dot{s}\equiv 0$). 
In terms of the adiabatic field, the momentum perturbation is given by
\begin{equation}
\delta q = -\dot{\sigma}\delta \sigma.
\end{equation}
The comoving density perturbation in terms of the 
adiabatic and entropy fields is\cite{bend-Gordon:2000hv,
wands-multi-omega}:
\begin{eqnarray}
\label{comoving-eq-simplest}
\epsilon_m &=&\delta X 
-\ddot{\sigma}\delta \sigma +[\delta V-V_{\sigma}\delta \sigma]
\nonumber\\
&\simeq& \delta X 
-\ddot{\sigma}\delta \sigma +V_{s}\delta s,
\end{eqnarray}
where the term related to the change of the basis of the adiabatic field
has been included in the definition of $\delta
X$\cite{entropy-V-Langlois:2006vv}.
At large scales the equation leads to
\begin{eqnarray}
\label{comoving-eq-simplest2}
\frac{\delta X}{X}
&=&\frac{\epsilon_m}{X} 
+\frac{\ddot{\sigma}}{X}\delta \sigma 
-\frac{V_{s}\delta s}{X}\nonumber\\
&\simeq & \frac{\ddot{\sigma}}{X}\delta \sigma 
-\frac{V_{s}\delta s}{X},
\end{eqnarray}
where the term proportional to $\frac{k^2}{a^2}\Psi$ has been disregarded.
Considering the perturbation of the inflaton velocity
$\frac{\delta(\dot{\sigma})^2}{\dot{\sigma}^2}=\frac{\delta X}{X}$ 
and the modulated
inflation scenario, $\dot{\zeta}_N$ is found to be
\begin{eqnarray}
\dot{\zeta}_N&\simeq& -H\frac{\delta X}{X}\nonumber\\ 
&\simeq& 
2\frac{V_{s}\delta s}{\dot{\sigma}^2},
\end{eqnarray}
where $\ddot{\sigma}\delta \sigma/X$ has been neglected. 
For ${\cal R}=-\delta N$, it is found that
\begin{equation}
\dot{\cal R} \simeq-2\frac{V_{s}\delta s}{\dot{\sigma}^2}.
\end{equation}
Introducing a bend parameter $\dot{\theta}\equiv -V_s/\dot{\sigma}$,
reveals\cite{bend-Gordon:2000hv}
\begin{equation}
\label{stan-rev}
\dot{\cal R} \simeq 
2H\frac{\dot{\theta}}{\dot{\sigma}}\delta s.
\end{equation}

In the above calculation the modulated inflaton velocity
$\delta(\dot{\sigma})=2\delta X$ has been
obtained directly from the comoving energy density $\epsilon_m$.
The calculation
in terms of the modulated inflation scenario is thus very simple and
straightforward compared with other calculations, which are
 based on the non-adiabatic pressure perturbation or the time-derivative
 of the comoving curvature perturbation on spatially flat slice.

The reason for $\dot{\cal R}\ne0$ is very clear in this scenario.
The constancy of the curvature perturbation is violated 
due to the inhomogeneities of the inflaton velocity
$\delta (\dot{\sigma}^2)\ne 0$, which is
 caused by the entropy field.
Then the inhomogeneities of the inflaton velocity 
$\delta(\dot{\sigma}^2)\ne 0$ causes inhomogeneities of the 
time spent during inflation, equivalently the
 inhomogeneities of the e-foldings $\delta N_{mod}\ne 0$ before
 the end of inflation. 
The result is also useful for estimating the second order corrections
from the potential.
Introducing the quadratic potential $\delta^{(2)}V=\frac{1}{2}m_s s^2$,
it is found that the second-order perturbation of the potential with
respect to the field $s$ leads to 
\begin{eqnarray}
{\cal R}^{(2)}&\simeq&-H\frac{m_s (\delta s)^2}{\dot{\sigma}^2}.
\end{eqnarray}

\subsection{Non-equilibrium correction 1}

The bend of the trajectory is important since the velocity perturbation
$\delta (\dot{\sigma})^2$ can become non-zero near the
bend, even if it arises late after the horizon exit.
At the bend of the trajectory, the velocity perturbation 
is given by
\begin{equation}
\label{decay-fl-vel}
\lim_{k\rightarrow 0} \delta(\dot{\sigma}^2) \simeq \dot{\theta}
\dot{\sigma}\delta s.
\end{equation}
In addition to the inhomogeneities that may appear at large scales,
there are small-scale (i.e., decaying) inhomogeneities of the inflaton velocity 
that can be related to the inhomogeneities in the slow-rolling velocity 
\begin{equation}  
\delta (\dot{\sigma}_{slow})\equiv\delta\left(-\frac{V_\sigma}{3H}
\right)\simeq -\frac{V_{\sigma ,s}}{3H}\delta s,
\end{equation}
which decays at large scales as $\propto k^2/a^2$\cite{Modulated1}.
The correction from such small-scale inhomogeneities can be dubbed
non-equilibrium corrections, which may become additional
source of the curvature perturbation.
More specifically, Eq. (\ref{decay-fl-vel}) suggests that 
$\delta (\dot{\sigma})^2$  soon decays to
reach $\delta(\dot{\sigma}^2) \simeq \dot{\theta}
\dot{\sigma}\delta s$ after
the horizon exit, and that at large scales the inflation velocity
perturbations do not lead to significant evolution $\dot{\cal R}\ne0$
 except for
the place where the inflation trajectory is curved.
The velocity perturbation may have a decaying
component accompanied by a significant decay factor $k^2/a^2\sim
e^{-2Ht}$\cite{Modulated1}. 
In what follows we show why the corrections added by the decaying
perturbations cannot be disregarded.

First, the small-scale inhomogeneities may affect the curvature
perturbations at least just after horizon crossing.
Then, we know that the perturbations added at small scales will be
frozen at large scales.
Therefore, if the small-scale corrections are significant, they can be
observed in the present Universe.  
Note that the factor $e^{-2Ht}$  in the integral with respect to the
time coordinate does not always lead to strong suppression in the
result. 
For the simplest example, the following integral is considered
\begin{equation}
\label{easy-eq}
\int^{t_e}_0 \delta C  H e^{-2Ht} dt \simeq \frac{1}{2}
\delta C,
\end{equation}
which is such that no exponential suppression remains after
integration. 

To show explicitly the significant effect from the decaying component
in the modulated inflation velocity, here a very simple mechanism is
considered to add a large non-Gaussian effect to the conventional
inflationary perturbation. 
Consider the standard
hybrid-type inflation with a non-standard interaction with addition
scalar fields $\chi_i$;
\begin{equation}
V_{int}\sim g^2 v_0 |\phi-\phi_{ESP}| \chi_i^2,
\end{equation}
where $v_0$ denotes an intermediate mass scale.
The inflaton potential during inflation is given by
\begin{equation}
V(\phi)=m_\phi^2\phi^2 + V_0.
\end{equation}
In this model the adiabatic inflaton field is $\phi$.
Enhanced symmetric point (ESP) at $\phi=\phi_{ESP}$ is defined as
the point where the scalar 
fields $\chi_i$ become (temporarily) massless during inflation. 
The inhomogeneities in $V_{int}$ caused by the light fields $\chi_i$
lead to the small-scale velocity perturbation 
\begin{eqnarray}
\delta\dot{\phi} &\simeq& 
\frac{ng^2 v_0 (\delta \chi)^2}{3H},
\end{eqnarray} 
where $n$ is the number of massless fields at the ESP.
Note that there is no bend in the classical (unperturbed) trajectory
in this model.
The small-scale inhomogeneities of the inflaton velocity thus
lead to the correction given by
\begin{equation}
\Delta R \equiv \int \dot{\cal R}dt 
\simeq 2\int H\frac{\delta \dot{\phi}}{\dot{\phi}}e^{-2H(t-t_{ESP})}dt,
\end{equation}
which can be very large when the inflaton passes through the ESP at
$t=t_{ESP}$. 
Considering the standard curvature perturbations 
${\cal R}_0 \sim H \delta \phi/\dot{\phi}$, we find
\begin{eqnarray}
\Delta R &\sim& \frac{\delta\dot{\phi}}{\dot{\phi}}\nonumber\\
 &\sim& \frac{ng^2 v_0 (\delta \chi)^2}{3H^2 \delta \phi}{\cal R}_0.
\end{eqnarray} 
It is useful to specify the level of non-Gaussianity by the non-linear
parameter $f_{NL}$,\footnote{See ref.\cite{Bartolo-text} for more
details.}
 which is usually defined by the Bardeen potential $\Phi$,
\begin{equation}
\Phi=\Phi_{Gaussian}+f_{NL}\Phi_{Gaussian}^2.
\end{equation}
Using the Bardeen potential, the curvature perturbation $\zeta$ is given
by
\begin{equation}
\Phi=\frac{3}{5}\zeta.
\end{equation}
When we consider ``additional'' non-Gaussianity created at the ESP,
the first-order perturbation is generated dominantly by the usual
inflaton perturbation $\delta \phi$.  
Therefore, the second-order perturbation is not
correlated to the first-order perturbation. 
In this case, $\zeta$ can be expanded by the $\delta N$ formalism as
\begin{equation}
\zeta \simeq N_\phi \delta \phi + N_\chi\delta\chi
+\frac{1}{2} N_{\phi\phi} \delta \phi^2 
+\frac{1}{2} N_{\chi\chi} \delta \chi^2 + ...,
\end{equation}
and we assume that the perturbation can be separated as
\begin{equation}
\zeta = \zeta^{(\phi)}+\zeta^{(\sigma)}.
\end{equation}
The calculation of the non-linear parameter $f_{NL}$ for the
uncorrelated perturbations $\delta \chi$ and $\delta \phi$ is discussed
in ref.\cite{Lyth_and_Rod_NG}, where a useful simplification 
for is 
\begin{equation}
f_{NL}\simeq \left(\frac{1}{1300}
\frac{N_{\sigma\sigma}}{N_\phi^2}\right)^3,
\end{equation}
where  $\delta \chi\sim\delta \phi$ is assumed for simplicity.
Therefore, the non-linear parameter for the present model is estimated
as
\begin{equation}
f_{NL}\sim \left(\frac{\Delta R}{1300{\cal R}_0^2}\right)^3
\sim \left(\frac{ng^2 v_0 }{10^3{\cal R}_0 H}\right)^3.
\end{equation}
The important suggestion from the model
 is that 
{\bf a significant scale-dependence may arise for the non-linear
parameter} 
 $f_{NL}(k)$ at a scale corresponding to $\phi=\phi_{ESP}$
 \cite{matsuda-SUSY}, where the $\chi_i$ fields become massless
at the horizon exit.
We show a typical situation in Fig. 1.
\begin{figure}[h]
 \begin{center}
\begin{picture}(400,240)(0,0)
\resizebox{12cm}{!}{\includegraphics{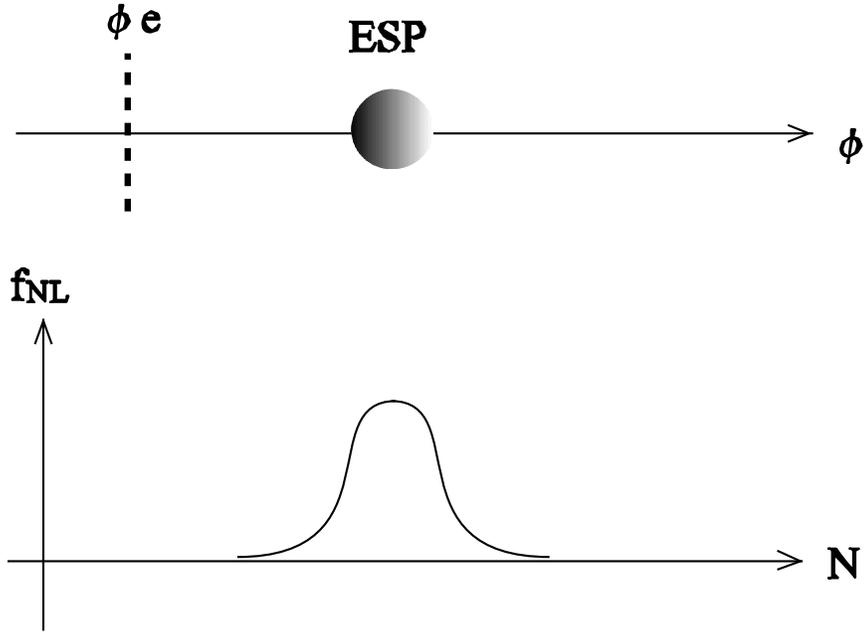}} 
\end{picture}
\label{fig:eps}
 \caption{In a brane inflation model one may find an enhanced symmetric
  point on the inflation trajectory, where massless excitations (open
  string modes) may arise\cite{matsuda-SUSY}. If the massless
  excitations are coupled to the 
  inflaton, the fluctuations of the massless mode may cause fluctuations
  of the inflaton velocity, which adds a significant non-Gaussian
  character to the spectrum.} 
 \end{center}
\end{figure}
Since ESPs may typically appear in a brane
Universe, there is a hope that we may scan the moduli or even find the
signature of extra dimensions from the consideration of the
scale-dependence in the cosmological perturbations
\cite{Kofman-modulated, roulette-inflation, matsuda-SUSY}.  

Again, what is important in this argument is that decaying-components
of the velocity perturbation may cause significant anomalies in the
spectrum, which may be seen in some non-linear parameter or in some
other cosmological parameters as a signature of massless excitations
during inflation.

Before closing this section, it would be useful to compare our model
with a model with a step (or a spike) in the potential\cite{Chen:2006xjb}.
A step in the potential leads to a time-dependent slow-roll parameters,
which causes $f_{NL}\ne 0$ for a single-field inflation model. 
In this case the origin of the higher order perturbations are terms
proportional to $\dot{\epsilon}$ or $\dot{\eta}$, which are not
important in the smooth potential but can cause significant
non-gaussianity at the step.
The obvious discrepancy between the model with step 
and our model with ESP is that in the former model the origin of the
non-gaussianity is $\dot{\epsilon}\ne0$ or $\dot{\eta}\ne 0$, while in
the latter model the origin is the second-order perturbation 
in the velocity.
Therefore, these two effects appear separately in the calculation and
give independent results for non-linear parameters.
In fact, in the present model $\dot{\epsilon}$ and $\dot{\eta}$ 
at the 0-th order are very small at the ESP.

\subsection{Non-equilibrium correction 2}
In addition to the non-equilibrium correction from the modulated
inflaton velocity, there can be significant correction from
$\ddot{\sigma}$ when deviation from slow-roll is 
significant\cite{stop-index, matsuda-stop-index}.
 $\ddot{\sigma}$ in the equation of $\dot{\cal R}$ is essential in
 explaining the evolution of ${\cal R}$ when the deviation is
 significant.
Allowing a short period of deviation from the slow-roll, the most
significant effect may occur when the inflaton temporarily stops.
Expressing the conventional curvature perturbation for slow-roll
inflation as ${\cal R}\simeq-\frac{H}{\dot{\sigma}}\delta \sigma$,
the divergence is brought about by $\dot{\sigma}\simeq 0$ in the
denominator. 
To understand the correction from $\ddot{\sigma}$ when the inflaton
stops, we split the inflaton velocity as 
\begin{equation}
\dot{\sigma} = \dot{\sigma}_s+\dot{\sigma}_d,
\end{equation}
where $\dot{\sigma}_s$ satisfies the slow-roll condition 
$\dot{\sigma}_s= -V_\sigma/3H$.
Considering the equation
\begin{equation}
\ddot{\sigma}+ 3H\left[\dot{\sigma}_s+\dot{\sigma}_d\right]
+V_\sigma=0,
\end{equation}
 the decaying velocity $\dot{\sigma}_d$ follows
$\ddot{\sigma}=-3H \dot{\sigma}_d$.
Replacing $\ddot{\sigma}$ in Eq.(\ref{comoving-eq-simplest}) by 
$-3H\dot{\sigma}_d$, it is found for $\dot{\theta}=0$;
\begin{eqnarray}
\dot{\cal R} &\simeq& 
-
6H \left(\frac{\dot{\sigma}_d}{\dot{\sigma}}\right)
\left(H\frac{\delta \sigma}{\dot{\sigma}}\right)\nonumber\\
&=&-
6H \epsilon_d^{-1} {\cal R}.
\end{eqnarray}
Here a parameter $\epsilon_d\equiv \dot{\sigma}/\dot{\sigma}_d$
is introduced, which leads to 
$\epsilon_d^{-1} \rightarrow \infty$ when the
 inflaton stops temporarily during inflation.
It is found from the equation that the curvature perturbation,
which may become anomalously large 
when the inflaton stops, will
decay rapidly as $\propto \exp[-3H\epsilon_d^{-1}t]$.
See also Ref.\cite{stop-index} for discussions in terms of the usual
 perturbation theory, and Ref.\cite{matsuda-stop-index} for 
another discussion in terms of the $\delta N$ formalism.

\section{Simple extension}

\underline{\bf The model}

Our second example is characterized by an extension of the inflation
kinetic term with a metric for the field space: 
\begin{equation}
G_{\phi \phi} = \omega^A(\phi,\chi), \,\,\, 
G_{\chi\chi}=\omega^B(\phi,\chi),
\end{equation}
where $\omega^A$ and $\omega^B$ are functions determined by the scalar fields.
For simplicity, the case with the diagonal metric is considered. 
The explicit form of the action is given by 
\begin{eqnarray}
\label{s-t-action}
S&=&\int d^4 x \sqrt{-g} \left[\frac{1}{2}M_p^2 {\cal R}
-\frac{\omega^A}{2}\left(\partial_\mu\phi\right)
\left(\partial^{\mu}\phi\right)\right.\nonumber \\ 
&&\left.-\frac{\omega^B}{2}\left(\partial_\mu \chi\right)
\left(\partial^{\mu} \chi\right)
-V(\phi,\chi)\right].
\end{eqnarray}
The field equations are
\begin{eqnarray}
\omega^A \ddot{\phi}+ \dot{\omega^A} \dot{\phi}
+ 3\omega^A H \dot{\phi} +V_\phi -\frac{1}{2}\omega^B_\phi\dot{\chi}^2
&=& 0\\
\omega^B \ddot{\chi}+ \dot{\omega^B} \dot{\chi}
+ 3\omega^B H \dot{\chi} +V_\chi - \frac{1}{2}\omega_\chi^A (\dot{\phi})^2
&=&0.
\end{eqnarray}
The comoving density perturbation is found to be
\begin{eqnarray}
\label{comoving-eq-simple}
\epsilon_m &\equiv& \delta \rho +3H G_{IJ}\dot{\phi}^J \delta \phi^I
\nonumber\\
&=&\delta X 
-(\omega^A \ddot{\phi}\delta \phi +\omega^B \ddot{\chi}\delta \chi )
+[\delta V - V_{\chi}\delta \chi -V_{\phi}\delta \phi]\nonumber\\
&& + \left(\frac{\omega^B_{\phi}}{2}\delta \phi(\dot{\chi})^2+
\frac{\omega^A_{\chi}}{2}\delta \chi(\dot{\phi})^2\right)
-\dot{\omega^A} \dot{\phi}\delta \phi- \dot{\omega^B} \dot{\chi}\delta \chi
\nonumber\\
 &\simeq& \delta X +\left(\frac{\omega^B_{\phi}}{2}\delta \phi(\dot{\chi})^2+
\frac{\omega^A_{\chi}}{2}\delta \chi(\dot{\phi})^2\right)
-\dot{\omega^A} \dot{\phi}\delta \phi- \dot{\omega^B} \dot{\chi}\delta \chi.
\end{eqnarray}
The adiabatic field in this model is defined by
$\dot{\sigma}^2/2\equiv X$.

\underline{\bf Calculation}

The velocity perturbation for the adiabatic field
$\dot{\sigma}^2 = \omega^A \dot{\phi}^2+\omega^B \dot{\chi}^2$ is given by
\begin{eqnarray}
\frac{\delta X}{X}&\simeq&
\frac{\epsilon_m 
- \left(\frac{\omega^B_{\phi}}{2}\delta \phi(\dot{\chi})^2+
\frac{\omega^A_{\chi}}{2}\delta \chi(\dot{\phi})^2\right)
+\dot{\omega^A} \dot{\phi}\delta \phi+ \dot{\omega^B} \dot{\chi}\delta \chi
}{X}.
\end{eqnarray}
Using the modulated inflation scenario, the form of 
$\dot{\cal R}$ in terms of the original fields $\phi$ and $\chi$ is found to be
\begin{equation}
\label{non-st-mat-original}
\dot{\cal R} \simeq 
H\left(\frac{\omega^B_{\phi}}{2X}\delta \phi(\dot{\chi})^2+
\frac{\omega^A_{\chi}}{2X}\delta \chi(\dot{\phi})^2\right)
-H\frac{\dot{\omega^A} \dot{\phi}\delta \phi+ 
\dot{\omega^B} \dot{\chi}\delta \chi}{X},
\end{equation}
which gives the evolution of the curvature perturbation in the
 slice defined for $\dot{\zeta}_N$.

In terms of the adiabatic field $\sigma$ and the entropy field $s$, 
the action is precisely the same as the model discussed in Sec.2.
Therefore, the evolution equation for the curvature perturbation is given by
\begin{equation}
\label{non-st-mat}
\dot{\cal R} \simeq -
2H\frac{V_{s}}{\dot{\sigma}^2}\delta {s},
\end{equation}
which is precisely the same as the two-field inflation model with 
the standard kinetic term.

\section{Modulated inflation for a generalized multi-field inflation}

\underline{\bf The model}

We consider multi-field inflation
with kinetic terms with a metric $G^{IJ}(\phi^L)$ in
field space.
The original action described by $\phi^I$ is;
\begin{equation}
S=\int d^4 x \sqrt{-g}\left[
\frac{R}{16\pi G}+P(X, \phi^I)
\right],
\end{equation}
where $X$ is given by
\begin{equation}
X\equiv -\frac{1}{2}G_{IJ}\partial_\mu \phi^I\partial^{\mu}\phi^J.
\end{equation}
In this section, a separation is considered
\begin{equation}
P(X, \phi^I)=K(X, \phi_I)-V(\phi^I)
\end{equation}
and set $8\pi G=1$ for simplicity.
In what follows we use the basic equations given in
Ref.\cite{Langlois:2008mn}.
The field equation is given by 
\begin{equation}
\ddot{\phi}^J 
+\left(
3H + \frac{\dot{G_{IJ}}}{G_{IJ}}+\frac{\dot{K_X}}{K_X}
\right)\dot{\phi}^J
-\frac{K_{[I]}-V_I}{K_X G_{IJ}}=0.
\end{equation}
The definition of the derivative $K_{[I]}$ may be somewhat confusing.
We introduce two different definitions for the derivatives
\begin{eqnarray}
K(X,\phi_I)_{[I]} &\equiv& K_X X_I -K_I\nonumber\\
K(X,\phi_I)_I  &\equiv& \frac{\partial K}{\partial \phi^I}.
\end{eqnarray}
The energy density for this action is given by
\begin{equation}
\rho = 2X K_X -K+V,
\end{equation}
which leads to the Friedman equation
\begin{equation}
H^2 = \frac{1}{3}\left(2X K_X-K+V\right)
\end{equation}
and the time-derivative of the Hubble parameter
\begin{equation}
\dot{H}=-X K_X.
\end{equation}
Combining the above equation with the continuity equation
$\dot{\rho}=-3H(\rho+p)$, it is found that 
\begin{equation}
\rho+p =-2\dot{H}.
\end{equation}

\underline{\bf Calculation}

{\bf For generalized multi-field inflation, the natural
definition of the adiabatic field $\tilde{\sigma}$ is 
\begin{equation}
\label{adi-st}
\dot{\tilde{\sigma}}^2 \equiv \rho+p = 2XK_X,
\end{equation}
which recovers the basic equation (\ref{orig}) in terms of 
$\tilde{\sigma}$.
The definition is very natural and consistent with the intrinsic property of the
adiabatic field.
However, in past studies a more simple definition $\dot{\sigma}\equiv
\sqrt{2X}$ has been considered.
The discrepancy caused by the definition of the adiabatic field may 
lead to an error in the result.}

In the followings, in addition to the natural definition of the
adiabatic field $\tilde{\sigma}$, the familiar definition
$\dot{\sigma}\equiv \sqrt{2X}$ is also considered  
so that these results from different definitions of the adiabatic field
can be compared.

The velocity perturbation for the adiabatic field
$\dot{\tilde{\sigma}}^2\equiv 2XK_X$ is given by
\begin{eqnarray}
\label{tilde-sigma-delta}
\frac{\delta(\dot{\tilde{\sigma}}^2)}{\dot{\tilde{\sigma}}}
&=&\frac{\delta X}{X}+\frac{\delta K_X}{K_X},
\end{eqnarray} 
whereas for the usual definition $\dot{\sigma}^2\equiv 2X$, 
it is found that
\begin{eqnarray}
\label{sigma-delta}
\frac{\delta(\dot{\sigma}^2)}{\dot{\sigma}^2} &=& \frac{\delta X}{X}.
\end{eqnarray}
The definition of the adiabatic field is important.
Using the adiabatic field and the $\delta N$ formalism,
 $\dot{\cal R}$ is calculated from Eq.(\ref{orig}) for the adiabatic
 field defined by $\tilde{\sigma}$, not by $\sigma$.
Obviously, the discrepancy
between Eqs.(\ref{tilde-sigma-delta}) and (\ref{sigma-delta}) is crucial
for the calculation.  
In order to calculate $\dot{\cal R}$
in terms of the $\delta N$ formalism,
the explicit form of $\delta X$ at large scales needs to be found.
As in the former arguments for simpler models in Sect.2 and 3, the
explicit 
form of the velocity perturbation
is obtained from the perturbation of the comoving energy density. 
The perturbation of the comoving energy density 
for the generalized multi-field inflation model is
\begin{eqnarray}
\epsilon_m &\equiv& 
\delta \rho + 3H K_X G_{IJ}\dot{\phi^J} \delta \phi^I,
\end{eqnarray}
where the energy density perturbation is given by
\begin{eqnarray}
\delta \rho &=& \delta X\left(K_X+2X K_{XX}\right)
+\delta \phi^I\left(2X K_{XI}-K_I+V_I\right)+ 
2X\delta^{(2)}K_X - \delta^{(2)}P
\nonumber\\ 
&\simeq& \frac{K_X}{c_s^2}\delta X +\delta \phi^I\left(2X K_{XI}-K_I+V_I
\right),
\end{eqnarray}
where the higher order perturbations $\delta^{(2)}K_X$ and
$\delta^{(2)}P$ has been disregarded in the last equation.
Here the effective sound speed defined by
$c_s^2 \equiv K_X/(K_X+2XK_{XX})$ is introduced.
Considering the expansion of the time-derivative, we find
\begin{eqnarray}
\dot{K_X}&=&K_{XX}\dot{X}+K_{XL}\dot{\phi}^L\nonumber\\
&=&K_{XX}\left(\frac{\dot{G_{IJ}}}{G_{IJ}}X+
\frac{1}{2}G_{IJ}\ddot{\phi}^I\dot{\phi}^J+
\frac{1}{2}G_{JI}\ddot{\phi}^J\dot{\phi}^I
\right)+K_{XL}\dot{\phi}^L,
\end{eqnarray}
which is used to rewrite the field equation as
\begin{equation}
\frac{1}{c_s^2}\ddot{\phi}^I+
\left[3H + \left(1+\frac{XK_{XX}}{K_X}\right)
\frac{\dot{G_{IJ}}}{G_{IJ}}+\frac{K_{XL}\dot{\phi}^L}{K_X}
\right]\dot{\phi}^I-\frac{G^{IJ}}{K_X}\left(K_{[J]}-V_J\right)=0.
\end{equation}
Using the field equation, the perturbation of the comoving energy
density is found to be given by
\begin{eqnarray}
\epsilon_m &\simeq& 
 \frac{K_X}{c_s^2}\delta X +\delta \phi^I\left(2X K_{XI}-K_I+V_I
\right) + 3H K_X G_{IJ}\dot{\phi^J} \delta \phi^I
\nonumber\\
&\simeq& 
 \frac{K_X}{c_s^2}\delta X +\delta \phi^I\left(
-K_I+K_{[I]}-\dot{G}_{IJ}\dot{\phi}^J
\left[K_X+XK_{XX}\right]\right).
\end{eqnarray}
We basically followed the calculation in Sec.2 and Sec.3.
In order to calculate the time-derivative of the metric for the field
space, the following expansion is considered\footnote{The simplification
is not valid when $G_{JL}=0$. The original equation must
be used for $G_{JL}=0$.}
\begin{eqnarray}
\dot{G}_{IJ}\dot{\phi}^J&=&\left[\partial_L G_{IJ}\dot{\phi}^L\right]
\dot{\phi}^J
\nonumber\\
&=&\frac{\partial_L G_{IJ}}{G_{JL}}2X,
\end{eqnarray}
which leads to $\epsilon_m$ given by
\begin{eqnarray}
\label{eps_m-original}
\epsilon_m &\simeq& 
 \frac{K_X}{c_s^2}\delta X +\delta \phi^I\left(
-K_I+K_{[I]}-\left(K_X+XK_{XX}\right)2X \frac{\partial_L G_{IJ}}{G_{JL}}
\right).
\end{eqnarray}
Here the difference between $K_{[I]}$ and $K_I$ is crucial for the model
discussed in Sec.3.

It would be useful to calculate $\dot{\cal R}$
from the usual definition of the adiabatic field $\sigma$ using the
modulated inflaton velocity.
Expressing the action in terms of the adiabatic and entropy fields, 
$G_{\sigma s}$ and $G_{ss}$ in the above equation must vanish.
The comoving density perturbation is thus given by\cite{Langlois:2008mn}
\begin{eqnarray}
\epsilon_m &\simeq& 
 \frac{K_X}{c_s^2}\delta X +\delta s\left(2X K_{Xs}-P_s
\right)+\delta \sigma\left(2X K_{X\sigma}-P_\sigma
\right) + 3H K_X G_{\sigma\sigma}\dot{\sigma} \delta \sigma\nonumber\\
&\simeq& 
 \frac{K_X}{c_s^2}\delta X +\delta s\left(
2X K_{Xs}-P_s\right)\nonumber\\
&&+ \delta \sigma\left[-P_\sigma+P_{[\sigma]}
-\left(K_X+XK_{XX}\right)2X 
\frac{\partial_\sigma G_{\sigma\sigma}}{G_{\sigma\sigma}}\right]
\nonumber\\
&&-(K_X+2XK_{XX})G_{\sigma\sigma} \ddot{\sigma} \delta \sigma\nonumber\\
&\simeq& 
 \frac{K_X}{c_s^2}\delta X +\delta s\left(
2X K_{Xs}-P_s\right),
\end{eqnarray}
where terms proportional to $\delta \sigma$ disappeared using 
$G_{\sigma\sigma}=1$ and $\ddot{\sigma}\simeq 0$.
To find the velocity perturbation, we need to find 
$\delta X$ from the comoving energy density.
The perturbation $\delta X$ caused by the
entropy perturbation $\delta s$ is given by 
\begin{eqnarray}
\frac{\delta X}{X} &\simeq& 
  -\frac{c_s^2}{XK_X}\left[
\left(2X K_{Xs}-P_s\right)\delta s\right].
\end{eqnarray}
where the adiabatic field is defined by 
$\dot{\sigma}\equiv \sqrt{2X}$.\footnote{With regard to 
our definition of $\delta X$, the change of the basis of
the adiabatic field is already included in the definition.
However, if one redefines the evolution of the curvature perturbation
by $\dot{\hat{\zeta}}_N \equiv 
- H \frac{\delta \dot{\sigma}}{\dot{\sigma}}
\equiv-H\left[\frac{\delta X}{2X}+\frac{\dot{e}^I_\sigma\delta
\phi^I}{\dot{\sigma}}
 \right]$,
where $\dot{e}^I_\sigma=P_s e^I_s/P_X\dot{\sigma}$ is the rate of change
of the adiabatic basis vector $e^I_\sigma$ in terms of the entropy basis 
vector $e^I_s$, it leads to
$\hat{\dot{{\cal R}}} \simeq 
  \frac{H}{2X P_X}\left[
(1+c_s^2)P_s\delta s-2Xc_s^2 K_{Xs}\delta s\right]$.
Compared with the result and the calculation presented in the previous
study, $\dot{\cal R}$ in Ref.\cite{Langlois:2008mn} is reprodued by
$\hat{\dot{{\cal R}}}$ redefined above.} 

Here the useful expression for $\dot{\cal R}$ is
written in terms of the original fields $\phi^I$ or the adiabatic field
defined by $\sigma$.
The calculation based on $\sigma$ is useful when $X$ has
important meaning in the string theory.
Our result is 
\begin{eqnarray}
\dot{\cal R}
&\simeq&H\frac{\delta (\dot{\tilde{\sigma}}^2)}{\dot{\tilde{\sigma}}^2}\nonumber\\
&=& \left[\frac{\delta X}{X} + \frac{\delta K_X}{K_X}\right]\nonumber\\
&=&H\left[\frac{\delta X}{X} + \frac{\delta X K_{XX} + 
\delta \phi^I K_{XI}}{K_X}\right]\nonumber\\
&=&H\left[\frac{\delta X}{X}\left(\frac{K_X+XK_{XX}}{K_X}\right) + 
\frac{\delta \phi^I K_{XI}}{K_X}\right]\nonumber\\
&\simeq&-H\frac{\tilde{c}_s^2\left(2X K_{Xs}-P_s\right)\delta s}{XK_X}
+H\frac{K_{Xs}}{K_{X}}\delta s,
\end{eqnarray}
where $\tilde{c}_s^2$ is defined by 
\begin{equation}
\tilde{c}_s^2 \equiv c_s^2\frac{K_X+X K_{XX}}{K_X} =
\frac{K_X+X K_{XX}}{K_X+2X K_{XX}}.
\end{equation}
This result is useful in practical situations where $\dot{\cal R}$ 
would be calculated in terms of $\sigma$.
For the original fields $\phi^I$, it can be found that
\begin{eqnarray}
\dot{\cal R}_{\tilde{\sigma}} &\equiv& 
H\frac{1}{XK_X}\left[
(K_X+XK_{XX})\delta X+XK_{XI}\delta \phi^I 
\right]\nonumber\\
&\simeq& 
\frac{\tilde{c}_s^2 H}{X K_X}\left[
(K_I-K_{[I]})\delta \phi^I
+(K_X+XK_{XX})2X \frac{\partial_L G_{IJ}}{G_{JL}}\delta \phi^I\right]
\nonumber\\
&&+H\frac{K_{XI}}{K_X}\delta \phi^I.
\end{eqnarray}

\section{Conclusions and discussions}
In this paper the time-dependence of the curvature perturbation is
considered in terms of the $\delta N$ formalism.
What is new in this study is
(1) the $\delta N$ calculation in terms of the modulated inflation
velocity 
(2) the explicit form of $\dot{\cal R}$ calculated with respect to the
adiabatic field $\tilde{\sigma}$. 
Our method is new and very simple, which can be used to understand more
exciting topics including the evolution during warm
inflation\cite{Matsuda-warm} and the evolution for the 
generalized gravity theory\cite{Matsuda-gen}.

Our last comment is for the importance of the non-equilibrium
corrections.
As we have discussed in this paper, there are many kinds of decaying
inhomogeneities that may not be disregarded.
For example, the inhomogeneities caused by the inflaton
velocity   
$\delta (\dot{\sigma}_{slow})\equiv\delta\left(-\frac{V_\sigma}{3H}
\right)$ is not significant at large scales, however
after time integration the small-scale inhomogeneities may leave
significant correction to the curvature perturbation, and the
correction will be
fixed at large scales due to the constancy of the curvature
perturbation.

In addition to the correction induced by the decaying inhomogeneities,
$\ddot{\sigma}\delta\sigma$ may 
lead to significant variation of ${\cal R}$.
Namely, one can find a significant result when the inflaton stops
 temporarily during inflation, where the curvature perturbation can be
 separated into (initially very large) decaying component and (standard)
 non-decaying components\cite{stop-index}. 
It is possible to extract the
non-decaying component of the curvature perturbation by using 
the $\delta N$ formalism \cite{matsuda-stop-index}; however the
evolution with deviation
is not clearly understood in more general situations.

The above two corrections may be dubbed non-equilibrium corrections,
and may be important in string cosmological models.
It is important to understand the evolution of the curvature
 perturbation in terms of the non-equilibrium corrections.

Our hope is to understand the effect of massless
excitations and non-equilibrium dynamics during inflation, so that we 
can find signatures of inflaton potential 
 and extra-dimensional structure in the sky.

\section{Acknowledgment}
We wish to thank K.Shima for encouragement, and our colleagues at
Tokyo University for their kind hospitality.

\end{document}